
\font\ninerm=cmr9   \font\twerm=cmr12

\baselineskip=6mm
\magnification 1200
\footline={\hfill}
\rightline{\bf{hep-ph 9403361}}
\rightline{SLAC-PUB-6464}
\rightline{Revised Version}
\rightline{May 1994}
\null\vskip .5truecm
\centerline{\twerm ANGULAR DISTRIBUTIONS IN THE DRELL-YAN PROCESS:}
\centerline{\twerm A CLOSER LOOK AT HIGHER TWIST EFFECTS}
\vskip 2truecm
\centerline{\twerm A. Brandenburg\footnote{$^{1}$}{\ninerm Max Kade
fellow}$^{,2}$,
~ S. J. Brodsky\footnote{$^{2}$}{\ninerm Work supported by the
Department of Energy, contract DE-AC03-76SF00515},
~ V. V. Khoze$^{2}$ ~ and D. M\"uller$^{2,}$\footnote{$^{3}$}{\ninerm Supported
by Deutscher Akademischer Austauschdienst}}
\vskip .5truecm
\centerline{\it Stanford Linear Accelerator Center}
\centerline{\it Stanford University, Stanford, California 94309}
\vskip 1.5truecm
\centerline{\twerm Abstract}
\noindent We calculate the angular distribution of the lepton produced in the
Drell-Yan reaction taking into account pion bound state effects.
We work in the kinematic
region where one of the pion constituents goes far off-shell,
which allows us to treat
the bound state problem perturbatively. We show that the angular
distribution is very sensitive to the shape of the pion
distribution amplitude. The model we discuss fits the data if we
choose a
two-humped pion distribution amplitude suggested by QCD sum rules.

\vskip 1truecm
\centerline{Submitted to {\it Phys. Rev. Let.}}

\vfil\eject
\pageno=1
\footline={\hss\tenrm\folio\hss}

Lepton pair production in hadron--hadron collisions provides a basic
testing ground for our understanding of strong interactions. Extensive
experimental and theoretical work has been done in the past two decades
(for reviews see Ref. [1]).
In particular, the angular distribution of the lepton pair has been studied
in detail, revealing a fatal disagreement of the QCD improved parton model
prediction [2,3] with the data [4-6].
Recently it has been proposed that this problem may be resolved if the
nontrivial structure of the QCD vacuum induces spin correlations
between the initial state partons [7]. In this letter we pursue another
way to go beyond the standard parton model picture, namely
we consider contributions
to the angular distribution induced by hadron bound state effects. Our approach
is close in spirit to the higher twist model of [8,9].
\par
The angular distribution of the $\mu^+$ in
$$ \pi^-+N \to \gamma^*+X\to \mu^++\mu^-+X \eqno(1)$$
may be parameterized in general as follows:
$$ {1\over \sigma}{d\sigma\over d\Omega}\sim
1+\lambda\cos^2\theta+\mu\sin 2\theta\cos\phi
+{\nu\over 2} \sin^2\theta\cos 2\phi. \eqno(2)$$
Here $\theta$ and $\phi$ are angles defined in the muon pair rest frame and
$\lambda$, $\mu$ and $\nu$ are angle--independent coefficients.
The naive parton model (Drell--Yan picture [10]) views the production of the
virtual photon $\gamma^*$ in (1) as originating from the annihilation of two
uncorrelated constituent quarks, resulting in an angular distribution of the
form $1+\cos^2\theta$.
This
result follows simply from the fact that the virtual photon is produced
transversely polarized in the annihilation of two on-shell fermions.
\par
In order to describe the boson transverse momentum distribution
$d^2\sigma/dQ_T^2$ one has to take into account radiative corrections
to the Drell--Yan model.
The $Q_T$-distribution has been calculated in the QCD--improved parton
model to the order of ${\cal O}(\alpha_s)$ with resummation of the soft gluons
at the
leading double logarithmic accuracy (see Ref. [11] and references
therein).
This approach was used in [3] to compute the angular distribution at
fixed transverse momentum. The deviations from the $1+\cos^2\theta$ behavior
were found to be less than $5\%$ in the range $0<Q_T<3$ GeV [3].
However, the NA-10
measurements from CERN [4] and the Chicago-Iowa-Princeton collaboration
[5,6] show
a quite different behavior.  In the limit where the momentum fraction $x$
of one of the pion constituents is very close to $1$ and for moderate
transverse momenta of the muon pair, the value of
$\lambda$ turns strongly negative [6], consistent with a $\sin^2 \theta$
distribution.
 This implies that in this kinematic limit
the virtual photon is produced with
longitudinal polarization, rather than transverse. Furthermore, the data
[4-6] is observed to have a strong azimuthal
modulation (nonzero $\mu$ and $\nu$ in (2)), an effect which is missing in
standard QCD.
The Lam--Tung sum rule [2],
$ 1-\lambda-2\nu=0,$
which follows from the approach used in [3] is also badly violated by the
experimental data.
\par
One way to go beyond the standard treatment is to take into account the pion
bound state effects [8,9]. We want to treat the bound state problem
perturbatively; thus we will restrict ourselves to a specific kinematic
region in which the momentum fraction $x$ of one of the pion constituents
is large, $x>0.5$.
  In fact, in
the large $x$ region the off-shell nature of the annihilating quark
from the projectile is crucial, and thus the operative subprocess must
involve the correlated multi-parton structure of the projectile.  In
effect the dominant subprocess in the off-shell domain is $\pi^- q \to
\mu^+\mu^- q.$
We resolve the pion by a single hard gluon exchange [12]. The main
contribution to reaction (1) then comes from the diagrams of Fig. 1a,b
[8,9].
We see from diagram 1a that the
 $\bar{u}$ quark propagator is far off-shell,
$p_{\bar{u}}^2=-Q_T^2/(1-x_{\bar{u}})$.
The
second diagram is required by gauge invariance\footnote{$^{\dag}$}{In a physical
gauge
the contribution of the second diagram is purely higher twist, that is
it contains extra powers of $\sqrt{Q_T^2/Q^2}$.}.
The leading contribution to the amplitude $M$ for the reaction
$$ u+\pi^- \to \gamma^*+X\to \mu^++\mu^-+X \eqno(3)$$
is obtained [12] by convoluting the partonic amplitude
 $T(u+\bar{u}d \to \gamma^*+d\to \mu^++\mu^-+d)$
 with the pion
distribution amplitude $\phi(z,\tilde{Q}^2)$,
$$ M=\int_0^1 dz \ \phi(z,\tilde{Q}^2) \ T,\eqno(4) $$
where $\tilde{Q}^2\sim Q_T^2/(1-x)$ is the cutoff for the integration
over soft momenta
in the definition of $\phi$. In the regime where $\tilde{Q}^2$ and
$Q^2$ are compatible, one cannot use the usual probabilistic
factorization of the structure functions and the hard annihilation
subprocess [13].
\par
For the hadronic differential cross section we have

$$ {Q^2d\sigma(\pi^-N\to\mu^+\mu^-X) \over dQ^2dQ_T^2dx_Ld\Omega}=
{1\over (2\pi)^4}{1\over 64} \int_0^1 dx_u G_{u/N}(x_u)\int_0^1 dx_{\bar{u}}
{x_{\bar{u}}\over 1-x_{\bar{u}}+Q_T^2/Q^2} |M|^2$$
$$\delta(x_L-x_{\bar{u}}+x_u-Q_T^2 s^{-1}(1-x_{\bar{u}})^{-1}) \
\delta(Q^2-sx_ux_{\bar{u}}+Q_T^2( 1-x_{\bar{u}})^{-1})
+\{u\to \bar{d},\bar{u}\to d\}.\eqno(5)$$
Here
$Q^{\mu}$ is the four-momentum of $\gamma^*$ in the hadronic center of mass
system,
$x_{u(\bar{u})}$ is the light-cone momentum
fraction of the $u(\bar{u})$ quark and $G_{u/N}$ is the parton distribution
function of the nucleon.
The longitudinal momentum fraction of the photon is defined as
$x_L=2Q_L/\sqrt{s}$ and it should be noted that its maximum value,
$x_L^{\rm max}=1 - s^{-1}(Q^2+2\ Q_T^2)$ is slightly less than $1$.
The second term on the right hand side of (5) is the
same as the first one with quark flavors interchanged. This term gives
the contribution from the nucleon sea.
In Fig. 1c we show a typical contribution
to the hadronic cross section.

We note that no primordial or intrinsic transverse momenta
have been introduced.
The single gluon exchange is the only source of $Q_T$  in the model discussed.
We also neglected the quark masses and the mass of the projectile which are
small
compared to $\tilde{Q}$.

In analogy to eq. (2) we parameterize the angular distribution as follows,
$$ {Q^2d\sigma \over dQ^2dQ_T^2dx_Ld\Omega}  \ \left(
{Q^2d\sigma \over dQ^2dQ_T^2dx_L }\right)^{-1}=$$
$$
{3 \over 4 \pi}{1 \over \lambda +3} (1+ \lambda \cos^2\theta
+\mu\sin 2\theta\cos\phi+{\nu\over 2} \sin^2\theta\cos 2\phi)
, \eqno(6)$$
where the angular distribution coefficients $\lambda$, $\mu$ and $\nu$
are now functions of the kinematic variables $x_L$,
$Q_T^2/Q^2$ and $Q^2/s$.

We work in the Gottfried-Jackson frame where the $\hat{z}$ axis is taken
to be the pion direction in the muon pair rest frame and the $\hat{y}$
axis is
orthogonal to the $\pi^{-} N$ plane.
With some algebra, using eqs. (4)-(6), we arrive at the following expression
for $\lambda$, $\mu$ and $\nu$,

$$ \lambda(\tilde{x},Q_T^2/Q^2)=2N^{-1}\bigg\{ \ (1-\tilde{x})^2 \left[
({\rm Im} \ I(\tilde{x}))^2+(F+{\rm Re} \ I(\tilde{x}))^2\right]$$
$$-4 Q_T^2/Q^2\tilde{x}^2F^2
+Q_T^4/Q^4\tilde{x}^2 F^2 \ \bigg\}, \eqno(7) $$
$$ \mu(\tilde{x},Q_T^2/Q^2)=-4 N^{-1} \sqrt{Q_T^2/Q^2} F \tilde{x}
 \bigg\{ \ (1-\tilde{x})\left[
 F+{\rm Re} \ I(\tilde{x}) \right]+Q_T^2/Q^2 \tilde{x} F \ \bigg\},
 \eqno(8) $$
 $$\nu(\tilde{x},Q_T^2/Q^2)= -8 N^{-1} Q_T^2/Q^2 \tilde{x} (1-\tilde{x}) F
\left[ F+{\rm Re}
 \ I(\tilde{x}) \right], \eqno(9) $$
where
$$ F=\int_0^1 dz {\phi(z,\tilde{Q}^2) \over z }, \eqno(10)$$
$$ I(\tilde{x})=\int_0^1 dz {\phi(z,\tilde{Q}^2) \over z \ (z+\tilde{x}
- 1 +i
\epsilon)},
\eqno(11) $$
and
 $$ N(\tilde{x},Q_T^2/Q^2)=2\bigg\{ \ (1-\tilde{x})^2 \left[
 ({\rm Im} \ I(\tilde{x}))^2+(F+{\rm Re} \ I(\tilde{x}))^2\right]$$
 $$+4 Q_T^2/Q^2\tilde{x}^2F^2
  +Q_T^4/Q^4\tilde{x}^2 F^2 \ \bigg\}. \eqno(12) $$
\noindent The variable $\tilde{x}$ acts to resolve the distribution amplitude
much like the Bjorken variable resolves the structure functions,
$$ \tilde{x} \equiv {x_{\bar{u}}\over 1+Q_T^2/Q^2}={1\over 2}
{x_L +\sqrt{x_L^2 +4s^{-1}(Q^2+Q_T^2)}
\over 1+Q_T^2/Q^2}
.\eqno(13)$$
\noindent The factors $1/z$  in eq. (10), (11) come from the gluon propagators
and the factors
$1/(z+\tilde{x}-1\pm i\epsilon)$ arise from the quark propagator of
Fig. 1b.

In contrast to Refs. [9] and [14] we did not omit terms
 ${\cal{O}}((1-x_{\bar{u}})Q_T^2/Q^2)$ and ${\cal{O}}((1-x_{\bar{u}})^{-1}
 \ Q_T^4/Q^4)$
 and of higher orders.
 The nucleon distribution function $G_{q/N}$ does not appear in
 (7)-(9);
 thus only the pion distribution amplitude $\phi(z,\tilde{Q}^2)$ has to be
specified.
 Before doing so we give some technical comments on our calculation.

 We note that the internal quark line of Fig. 1b
 can go on-shell.
 The amplitude $M$ of equation (4), however, is always regular due to
 the $z$-integration [14] for realistic choices of $\phi(z,\tilde{Q}^2)$.
 This also can be read off from (11).
 The fact that the internal line goes on-shell
 does not cause a Sudakov suppression
 since our diagrams are the lowest order contribution
 of an {\it inclusive}  process. In other words gluon emission to the final
 state will occur in the higher order corrections. Only when $x_{\bar{u}}$
 approaches unity, where gluon emission is prohibited by kinematics,
 the Sudakov suppression
 will arise.

 Our model and the parton model are not complementary, but rather
 different approximations to the Drell-Yan process.
 The diagrams of Fig. 1a,b give the {\it whole} leading order contribution
 in the specific kinematic region of large enough $ x_{\bar{u}}$,
 $x_{\bar{u}} > 0.5$ [9].
 This is so because the gluon exchange is the resolution of the pion
 bound state and not a radiative correction.

 Now we can present our final results for $\lambda$, $\mu$ and
 $\nu$ for different choices of the pion distribution amplitude
 $\phi(z,\tilde{Q}^2)$. We find in general that the values of $\mu$ and $\nu$
 are very sensitive to the choice of $\phi(z,\tilde{Q}^2)$
 which we always take to be positive, symmetric,  i.e.
 $\phi(z,\tilde{Q}^2)=\phi(1-z,\tilde{Q}^2)$, and normalized, $\int_0^1 dz
 \ \phi(z,\tilde{Q}^2)=1$. Thus we will not restrict
 ourselves to the simplest case of $\phi(z) \sim \delta (z-1/2)$ considered
 in [9].

 This sensitivity can be illustrated for the special case of $\tilde{x}=0.5$
 for which ${\rm Re} \ I \ = \ -2F$.
 From eqs. (7)-(9) we get,
 $$\lambda (\tilde{x}=0.5, Q_T^2/Q^2) =
 { 1+4\pi^2 a^2 -4Q_T^2/Q^2 +Q_T^4/Q^4 \over 1+4\pi^2 a^2 +4Q_T^2/Q^2
+Q_T^4/Q^4}
 ,\eqno(14)$$
 $$\mu (\tilde{x}=0.5, Q_T^2/Q^2) =
 { 2\  \sqrt{Q_T^2/Q^2}\ (1-Q_T^2/Q^2)\over 1+4\pi^2 a^2 +4Q_T^2/Q^2 +Q_T^4/Q^4}
 ,\eqno(15)$$
 $$\nu (\tilde{x}=0.5, Q_T^2/Q^2) =
 { 4\  Q_T^2/Q^2\over 1+4\pi^2 a^2 +4Q_T^2/Q^2 +Q_T^4/Q^4}
 ,\eqno(16)$$
 where
 $a\equiv\phi (z=0.5)/F.$

 From these formulas we see that $\mu$ and $\nu$ in this case are not suppressed
 only if $a$ is a sufficiently small number. This is true for the so-called
 two-humped distribution amplitude [15] which has a dip around $z=0.5$.
 On the other hand, the choice of a convex distribution amplitude, e.g. the
asymptotic
 one, $\phi(z)=6 z (1-z)$, will always produce suppressed $\mu$ and $\nu$
 at $\tilde{x}=0.5$.

 We return now to our general results, eqs. (7)-(9).
 In Fig. 2 we plot $\lambda$, $\mu$, $\nu$ and $2\nu-(1-\lambda)$ versus
 $x_{\bar{u}}$
 for $\sqrt{Q_T^2/Q^2}=0.25$ for different choices of $\phi(z,\tilde{Q}^2)$
 together with the data of Ref. [5].
 For the two-humped distribution amplitude we have chosen the evolution
 parameter $\tilde{Q}^2$ to be effectively $\sim 4 \ {\rm GeV}^2$.
 The solid line is the result for the two-humped
 $\phi(z)$
 where powers of $(Q_T^2/Q^2)^{n/2}$ were dropped for $n \ge 3$ in eqs.
 (7)-(9).
 We note that corrections
 to our model may induce such terms, thus the difference between the dashed and
the solid lines
 should be viewed as the uncertainty of our predictions. We also show the data
points
 of Ref. [5] averaged in the intervals $4.05 < \sqrt{Q^2} < 8.55 \ {\rm GeV}$
and
 $0 < \sqrt{Q_T^2} < 5 \ {\rm GeV}$.

 In Fig. 3 the same quantities are shown versus $\sqrt{Q_T^2}$ ~ for
$x_{\bar{u}}=0.6$
 and $\sqrt{Q^2}=6 \ {\rm GeV}$. The data points in this case are averaged over
intervals
 $4.05 < \sqrt{Q^2} < 8.55 \ {\rm GeV}$ and $0.2<x_{\bar{u}}< 1$ and
 taken from Ref. [5].
 All the data points were averaged over the intervals defined above in Ref. [5].
 We would rather prefer to use the unaveraged data which are not  available.
 The use of the averaged over $x_{\bar{u}}$ data in Fig. 3 forced us to fix
 the value $x_{\bar{u}}=0.6$ for our theoretical prediction which is rather
 low for our model and pushes it to the limits of its applicability.

 Finally  we would like to
 comment on the limitations of our model
 and corrections to it.
 The bound state effects considered here should have received a truly
non-perturbative
 treatment. We have restricted ourselves to a perturbative approximation to the
problem.
 This approximation makes sense only at large enough $x$ which we have chosen to
be
 $>0.5$. The contribution of more than one hard gluon exchange will be
suppressed
 by powers of $\alpha_s$ in this case.
 The contribution of soft gluons to the  pion bound state is taken into
 account in the evolution of the distribution amplitude.
 The higher Fock states of the pion
 are expected to be suppressed when $x$ is large enough [12].
 The pion and the nucleon are not treated symmetrically in our model,
 namely the nucleon bound state effects are not taken into account since
 in the kinematic region we consider, $x_{\bar{u}}$ is always large and
 $x_u$ is always small.
 Gluon emission to the final state will first contribute to the evolution
 of the parton distribution functions. Hard gluon emission is suppressed by
powers of $\alpha_s$.
 No attempt of a systematic inclusion of higher order or mass effects was made.

The coefficient functions
$\lambda$, $\mu$, and $\nu$ at large $x >
0.5$ are very sensitive to the shape of the projectile's distribution
amplitude $\phi(z,\tilde{Q}^2)$, the basic hadron wavefunction which describes
the distribution of light-cone momentum fractions in the lowest-particle
number valence Fock state. Measurements of
meson form factors [12]  and other exclusive and semiexclusive
 processes [16] at large momentum transfer
can only provide
global constraints on the shape of
$\phi(z,\tilde{Q}^2)$; in contrast, the angular dependence of the lepton pair
distributions can be used to provide local measurements of the shapes of
these hadron wavefunctions.
Detailed measurements of the angular distribution of
leptons as a function of both $x$ and $Q_T$ for the reactions $H
p \to \ell^+ \ell^- X$ for the whole range of fixed target beams $H =
\pi, K, \bar p, p,$ and $n$ will open up a new window on the
structure of hadrons at the amplitude level.

Our analysis shows that the broad, two-humped, distribution amplitude for
the pion which was obtained within the context of
 QCD sum rules [15] can account for the main features
of the data.  In contrast,  narrow momentum distributions,
characteristic of
weak hadronic binding, predict the wrong sign for the observed azimuthal
angular coefficients $\mu$ and $\nu.$

\bigskip
{\bf Acknowledgments}

\noindent We wish to thank V. Telegdi for conversations regarding
the experimental data, particularly the fact that the observed
increase of the variable $\nu$ with the dimuon transverse momentum
is at variance with perturbative QCD calculations.
We are grateful to Yu. L. Dokshitzer, O. Nachtmann and W. J.
Stirling for useful suggestions and comments
on the manuscript.
We also benefited from discussions with H. Anlauf, V. M. Braun, T. Hyer,
 V. A. Khoze,
and M. Peskin.

 \vfil\eject

 \centerline{\twerm References}

 \item{[1]}K. Freudenreich, {\it Int. J. Mod. Phys.} {\bf A 5}(1990) 3643;
  Yu.L. Dokshitzer, D.I. Dyakonov and S.I. Troyan,
 {\it Phys. Rep.} {\bf 58 } (1980) 269;
  Proceedings of the Workshop on Drell--Yan Processes,
 Fermilab, Batavia, 1982
 \item{[2]} C.S. Lam and W.K. Tung, {\it Phys. Rev.} {\bf D 21} (1980) 2712
 \item{[3]} P. Chiappetta and M. Le Bellac, {\it Z. Phys.} {\bf C 32} (1986) 521
 \item{[4]} NA10 Collab. S. Falciano et al., {\it Z. Phys.} {\bf C 31} (1986)
513;
  NA10 Collab. M. Guanziroli et al., {\it Z. Phys.} {\bf C 37} (1988) 545
 \item{[5]} J.S. Conway et al., {\it Phys. Rev.} {\bf D 39} (1989) 92
 \item{[6]}J.G. Heinrich et al., {\it Phys. Rev.} {\bf D 44} (1991) 44
   \item{[7]} A. Brandenburg, O. Nachtmann and E. Mirkes, {\it Z. Phys.} {\bf C
60}
  (1993)  697
 \item{[8]} E.L. Berger and S.J. Brodsky, {\it Phys. Rev. Lett.}
  {\bf 42} (1979) 940;
    S.J. Brodsky, E.L. Berger and G.P. Lepage in
  Proceedings of the Workshop on Drell--Yan Processes,
 Fermilab, Batavia, 1982, p. 187
 \item{[9]} E.L. Berger, {\it Z. Phys.} {\bf C 4} (1980) 289
 \item{[10]} S.D. Drell and T.M. Yan, {\it Phys. Rev. Lett.} {\bf 25} (1970) 316
 \item{[11]} G. Altarelli, R.K. Ellis, M. Greco and G. Martinelli,
  {\it Nucl. Phys.}{\bf B 246} (1984) 12
  \item{[12]} G.P. Lepage and S.J. Brodsky, {\it Phys. Rev.} {\bf D 22} (1980)
2157
  \item{[13]} S.J. Brodsky, P. Hoyer, A.H. Mueller and W.K. Tang, {\it
  Nucl. Phys.} {\bf B 369} (1992) 519; K. Eshola, P. Hoyer, M. Vanttinen, R.
Vogt (to be
 published)
  \item{[14]} S. Matsuda, {\it Phys. Lett.} {\bf B 119} (1982) 207
  \item{[15]}V.L. Chernyak and A.R. Zhitnitsky,
  {\it Phys. Rep.} {\bf 112} (1984) 173
  \item{[16]} A.V. Efremov and A.V. Radyushkin, {\it Phys. Lett.} {\bf
  B 94} (1980) 245;
   M.K. Chase, {\it Nucl. Phys.} {\bf B 167} (1980) 125;
   S.J. Brodsky and G.P. Lepage, {\it Phys. Rev.} {\bf D 24}
  (1981) 1808;
   E. Braaten, {\it Phys. Rev.} {\bf D 28} (1983) 524;
   E. Maina and G. Farrar, {\it Phys. Lett.} {\bf B 206} (1988)
  120;
   T. Hyer, {\it Phys. Rev.} {\bf D 48} (1993) 147
  \vfil\eject
  \centerline{\twerm Figure Captions}
  \vskip 1truecm

  \item{\rm Fig. 1:}Diagrams (a) and (b) give the leading contribution
  to the amplitude of reaction (4). Diagram (c) gives a typical (one
  out of four) contribution to the cross section (6).

  \bigskip

  \item{\rm Fig. 2:}The angular distribution coefficients $\lambda$,
  $\mu$ and $\nu$ and the Lam--Tung combination,
  $2\nu-(1-\lambda)$, in the Gottfried-Jackson frame, ~ versus
 $x_{\bar{u}}$ for $\sqrt{Q_T^2/Q^2}=0.25$.
 The dotted line corresponds to $\phi(z)=\delta (z-1/2)$,
 the dashed-dotted line corresponds to the asymptotic
 $\phi(z)=6z(1-z)$ and the dashed line shows the results for
 the two humped distribution amplitude, $\phi(z)=26z(1-z)(1-50/13 \ z(1-z))$.
 The solid line is the result for the two-humped
 $\phi(z)$ where powers of $(Q_T^2/Q^2)^{n/2}$ were dropped for $n \ge 3$ in
eqs.
 (13)-(15). The data points (averaged as explained in the text) are taken from
Ref. [5].
 \bigskip

  \item{\rm Fig. 3:}The same quantities as in Fig. 2 are shown.
  versus $\sqrt{Q_T^2} \ $ for $\ x_{\bar{u}}=0.6$ ~ and ~ ~ ~ ~ ~ ~
  $\sqrt{Q^2}=6$ GeV.

   \end